\documentclass{article}

\usepackage{arxiv}

\usepackage[utf8]{inputenc} % allow utf-8 input
\usepackage[T1]{fontenc}    % use 8-bit T1 fonts
\usepackage{hyperref}       % hyperlinks
\usepackage{url}            % simple URL typesetting
\usepackage{booktabs}       % professional-quality tables
\usepackage{amsfonts}       % blackboard math symbols
\usepackage{nicefrac}       % compact symbols for 1/2, etc.
\usepackage{microtype}      % microtypography
\usepackage{lipsum}		% Can be removed after putting your text content
\usepackage{graphicx}
\usepackage[numbers]{natbib}
\usepackage{doi}

\usepackage{amsmath}
\usepackage{booktabs}
\usepackage{multirow}
\usepackage{amsfonts}

\usepackage{microtype}
\usepackage{soul}
\usepackage[inline]{enumitem}
\usepackage{soul}
\usepackage{subcaption}
\usepackage{url}
\newcommand{\todo}{\textcolor{red}}

\usepackage{graphicx}
\usepackage{subcaption}
\usepackage{algorithm2e}
\usepackage{color}
\usepackage{tikz}
\usetikzlibrary{calc}

\newcommand{\Hcal}{\mathcal{H}}
\newcommand{\Hcalt}{\mathcal{H}_t}

\newcommand{\Ucal}{\mathcal{U}}

\newcommand{\deltaequal}{\overset{\Delta}{=}}

\def\eqref#1{equation~\ref{#1}}
\def\1{\bm{1}}

\DeclareMathAlphabet{\mathsfit}{\encodingdefault}{\sfdefault}{m}{sl}
\SetMathAlphabet{\mathsfit}{bold}{\encodingdefault}{\sfdefault}{bx}{n}

\def\Hcal{{\mathcal{H}}}

\def\Ucal{{\mathcal{U}}}

\title{New Perspectives on the Evaluation of Link Prediction Algorithms for Dynamic Graphs}

\author{Raphaël Romero\\
Ghent University\\
raphael.romero@ugent.be
\And
Tijl De Bie\\
Ghent University\\
tijl.debie@ugent.be
\And
Jefrey Lijffijt\\
Ghent University\\
jefrey.lijffijt@ugent.be}
\hypersetup{
pdftitle={New Perspectives on the Evaluation of Link Prediction Algorithms for Dynamic Graphs},
pdfauthor={Raphaël Romero, Jefrey Lijffijt, Tijl De Bie},
pdfkeywords={Temporal Networks  \and Visualization \and Link Prediction},
}

\begin{document}
\maketitle

\begin{abstract}
	There is a fast-growing body of research on predicting future links in dynamic networks, with many new algorithms. Some benchmark data exists, and performance evaluations commonly rely on comparing the scores of observed network events (positives) with those of randomly generated ones (negatives). These evaluation measures depend on both the predictive ability of the model and, crucially, the type of negative samples used. Besides, as generally the case with temporal data, prediction quality may vary over time. This creates a complex evaluation space.
    
	In this work, we catalog the possibilities for negative sampling and introduce novel visualization methods that can yield insight into prediction performance and the dynamics of temporal networks. We leverage these visualization tools to investigate the effect of negative sampling on the predictive performance, at the node and edge level. We validate empirically, on datasets extracted from recent benchmarks that the error is typically not evenly distributed across different data segments. Finally, we argue that such visualization tools can serve as powerful guides to evaluate dynamic link prediction methods at different levels. 
\end{abstract}

% keywords can be removed

\newcommand{\addref}{\todo{reference here}}

\section{Introduction}

Many real-world phenomena such as biological ecosystems, networks of computers, email exchanges, and social interactions can be understood as a set of nodes interacting through time \cite{holmeTemporalNetworksModeling2021}. This type of data is commonly represented as a dynamic graph. While initial attempts at modeling the temporal evolution of networks typically aggregated the interactions into a sequence of static graphs, recent efforts consider how to directly model time continuously (i.e., without aggregation), preventing the loss of some fine-grained temporal information during data preprocessing.

The resulting type of data is commonly framed \emph{Continuous-Time Dynamic Graphs} (CTDGs). 
Learning from such CTDGs has recently become an active field of research, as suggested by recent benchmarks \cite{huangTemporalGraphBenchmark2023}. A crucial task thereof is Dynamic Link Prediction (DLP), where the goal is to predict future links from a history of observed ones. Successfully predicting interactions in a nearby future can have numerous applications, such as recommending items on online platforms, or predicting the evolution of epidemics through time. 

However, the DLP task has faced a reproducibility crisis akin to the one observed in static link prediction \cite{poursafaeiBetterEvaluationDynamic2023}. Indeed, while typical DLP evaluation pipelines merely compared the scores of query events to the scores of competing negative events selected at random, recent work has emphasized that the way these negative events are selected can have a dramatic impact on DLP performance measures, to the point that some sophisticated methods can get outperformed by trivial, parameter free heuristics \cite{poursafaeiBetterEvaluationDynamic2023}.
Thus, the awareness is growing that  dynamic link prediction performance measures not only depend on the model quality, but crucially also on the strategy for sampling negative events.

\begin{figure}[t]
	\includegraphics[width=\linewidth]{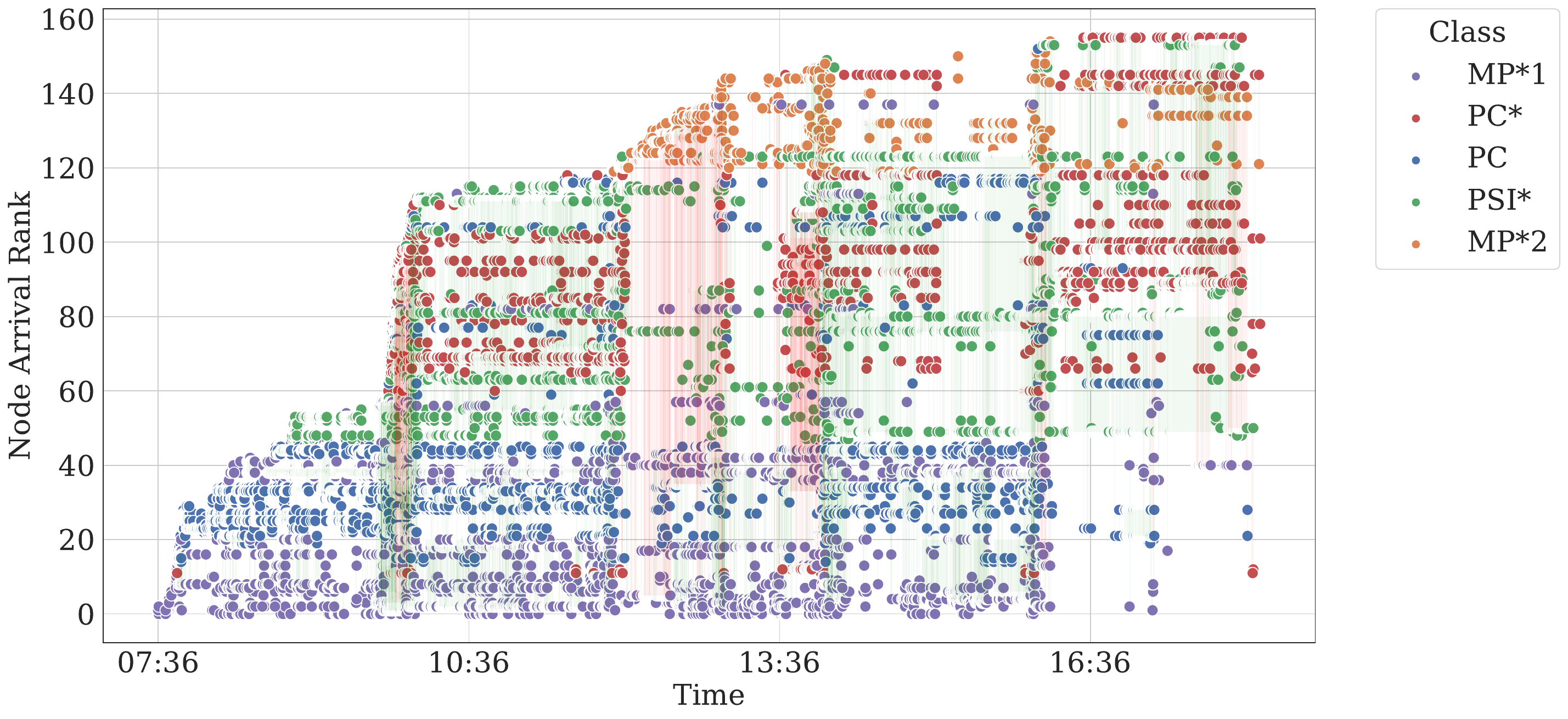}
	\caption{Example of a Temporal Node Activity plot on the High School dataset.
		The nodes are colored by class that the student is taking. The edges are colored green if they connect two nodes from the same class and red otherwise.
		Some structural information can be observed, such as the fact that the inter-class edges are more likely to occur during break times (10 am, 12--2 pm, 4 pm).\label{fig:tnt_highschool}}
\end{figure}

Furthermore, unlike conventional machine learning tasks targeting independent and identically distributed data, prediction errors in DLP are not distributed uniformly over the data. Hence, the use of visualization emerges as a vital tool for effectively charting a model's performance across different data segments.
However, in contrast to static graphs, Dynamic Graphs are notoriously hard to visualize \cite{holmeTemporalNetworkTheory2019}.  
Longitudinal visualization, while inherently limited \cite{linharesVisualisationStructureProcesses2019}, has not yet been applied to visualize predictive performance of DLP algorithms.
However, as seen in Figure $\ref{fig:tnt_highschool}$, visualizations such as Temporal Node Activity plots \emph{can}, to some extent, help to reveal some aspects of the temporal structure of CTDGs. 
In this paper, we aim to explore the opportunities visualization techniques offer in a predictive context.
Specifically, we propose the following contributions.

\textbf{Contributions:}
\begin{itemize}[topsep=0pt]
  \item \emph{Negative sampling} (Section 3): We explore the diversity of negative sampling strategies. To this aim, \ul{we propose a taxonomy separating negative sampling strategies into ten categories, and show their use in evaluation leads to different insights on the capabilities of a DLP algorithm}. Through careful evaluation of a state-of-the-art algorithm, we validate empirically the sensitivity of the performance (measured by AUC) to negative sampling.
  \item \emph{Visualization} (Section 4): \ul{We propose simple and effective lossless visualization techniques called the Temporal Node Activity and Temporal Edge Activity plot}, enabling the visualization the activity of nodes and edges over time without the need for aggregation.
%  The construction of these visualizations is described in subsection \ref{sec:visualizations}.
  Finally, \ul{we illustrate two ways in which the proposed visualization tools can yield more detailed insights into the performance of DLP algorithms}. 
\end{itemize}

\noindent
An open source implementation of the method and experiments is made available at \url{https://github.com/aida-ugent/dlp_viz}.

\section{Link Prediction on Dynamic Graphs}
\label{background}
A \textbf{Continuous-Time Dynamic Graph} (CTDG) is defined as a time-ordered stream of interactions $\mathcal{H} = \{(u_m, v_m, t_m)\ |\ m = 1, \ldots, M\}$, where for the $m$-th event, $u_m$ and $v_m$ are source and destination nodes belonging to a set of nodes $\mathcal{U}$, and $0\leq t_1 \leq t_2 \leq \ldots \leq t_M$ are the timestamps of interactions.
A \textbf{DLP algorithm} is a function that computes a score $s(u,v,t\ |\ \mathcal{H}_t)$ for any event $(u,v,t)$ \emph{conditioned} on the past events denoted $\mathcal{H}_t \deltaequal \{(u_m,v_m,t_m)\in\mathcal{H}\ |\ t_m<t\}$. DLP algorithms are typically \emph{trained} on a training history $\mathcal{H}_{\text{train}} \deltaequal \mathcal{H}_{t<t_{\text{train}}}$ and evaluated on a test history $\mathcal{H}_{\text{test}} \deltaequal \mathcal{H}_{t\geq t_{\text{train}}}$. The training set can be further split into train and validation, with typical proportions of 70\%, 15\%, and 15\% for the train, validation, and test sets, respectively.

\begin{figure}[t]
  \centering
  \includegraphics[width=\linewidth]{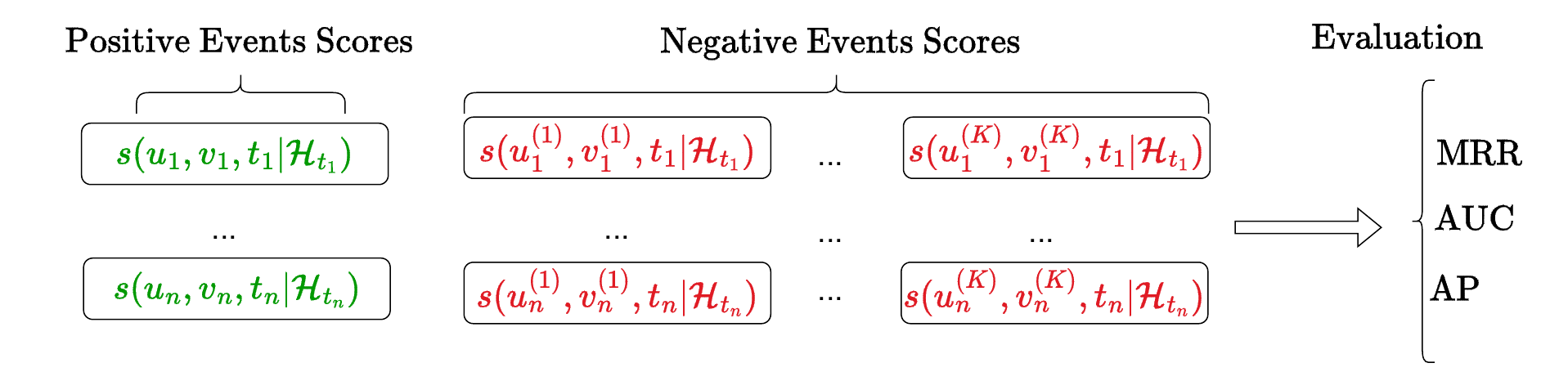}
  \caption{A typical evaluation pipeline for Dynamic Link Prediction: the score of each positive event is calculated along with the scores of $K$ negative events.
  The resulting scores are then used to compute the AUC, MRR and AP measures.
  The way the negative edges $(u_i^{(k)}, v_i^{(k)})$ are selected is a crucial component of the evaluation, as it can have a dramatic impact on the performance measures.}
  \label{fig:dlp_eval}
\end{figure}

DLP algorithms are typically \emph{evaluated} by comparing the scores of \emph{positive} events $s(u,v,t\ |\mathcal{H}_t)$ with the scores $s(u',v',t\ |\mathcal{H}_t)$ of competing events $(u',v',t)$ occurring at time $t$ and involving an edge $(u',v')$ other than the positive one. In theory, a perfect model should be able to rank the positive event $(u,v,t)$ first in the list of \emph{all possible events} at time $t$. However, generating this list of all possible events, let alone calculating a score for each of them, is often intractable. Instead, most evaluation methods select $K$ negative events at random and compare the scores of the positive events with those of the negative events. This setup is summarized in Figure \ref{fig:dlp_eval}. Crucially, the \emph{procedure} used to generate the negative samples can have a dramatic impact on the resulting Link Prediction measure.

DLP algorithms can be split into several main categories:
\begin{itemize}[topsep=0pt]
	\item \emph{Heuristic approaches} consist in calculating a statistic for each edge $(u,v)$ at time $t$, summarizing the historical information available for predicting that edge, up to that time $t$. 
In this paper, we will consider two simple instances of these heuristic approaches. 
\textbf{EdgeBank} (introduced in \cite{poursafaeiBetterEvaluationDynamic2023}) predicts $s(u,v,t\ |\Hcal_t)=1$ if the edge $(u,v)$ has an interaction previous to t. 
\textbf{Preferential Attachment (PA)} (first described in \cite{LibenLP2007}) scores events $(u,v,t)$ as the product of their temporal degrees $deg_u(t)\times deg_v(t)$, where $\deg_u(t)$ is the number of events $(u',v',t)$ such that $u'=u$ or $v'=u$. 
While simple, these baselines provide good reference points to assess the performance of more sophisticated methods, as they provide a \emph{controlled source} of false negatives and false positive. 
Indeed, Edgebank produces false negatives whenever the edge $u,v$ involved in the event $(u,v,t)$, and false positives whenever the edge involved in the negative event is historical. Similarly, PA produces false negatives whenever any of the nodes involved in the positive event is inductive, and false positives whenever both the nodes involved in the negative event are historical.
\item Recent efforts, however, focused on \emph{Dynamic Node Representation Learning}. As surveyed in \cite{kazemiRepresentationLearningDynamic2020a}, these methods can be described by an encoder-decoder architecture. 
% \newcommand{\dec}[1]{\textbf{DEC}_{\phi}\left(#1\right)}
% \newcommand{\enc}[1]{\textbf{ENC}_{\psi}(#1)}
% $s(u,v,t|\mathcal{H}_t) = \dec{\enc{u,t|\Hcal_t}, \enc{v,t,\Hcal_t}}$
The encoder is a function that maps each node $u$ to a time-varying representation $h_u(t|\mathcal{H}_t)$, while the decoder is a function that maps a pair of nodes $(u,v)$ and a time $t$ to a score $s(u,v,t|\mathcal{H}_t)$.
This approach extends concepts from the extensive literature on static graph embedding, where the objective is to learn, for each node, a representation that summarizes all relevant structural information at the node level. A similar concept is applied in the context of dynamic graphs, where the key distinction lies in allowing the representation to evolve over time as more events are observed. In the remaining of this paper we will consider the \textbf{TGN-attn} method, an instance of the Temporal Graph Network \cite{Rossi_Chamberlain_Frasca_Eynard_Monti_Bronstein_2020} that combines graph attention and gated recurrent units into a single temporal graph embedding method.  
\end{itemize}

% \subsection*{Temporal Edge Traffic Plots}

% Temporal Edge Traffic plots, introduced in \cite{poursafaeiBetterEvaluationDynamic2023} are 2-d scatter plots representing the temporal activity of edges (i.e. node-pairs) over time. 
% They are defined as follows. For each edge $(u,v)$ involved the CTDG we compute its first and last activation time $t_{min}(u,v)$ and $t_{max}(u,v)$. Next, the edges are sorted by ascending first activation time $t_{min}(u,v)$. In case of ties, the edges are sorted by ascending $t_{max}(u,v)$.  This yields for each edge a rank $r(e)$. The $x$-coordinate $x(u,v)\in [0,1]$ is calculated by normalizing the rank of edge $e$ by the total number of edges in the CTDG.
% Finally, each event $(u,v,t)$ gets represented by a point at coordinates $(x(u,v), t)$.
% These plots allow visualizing the proportion of edges seen during training/testing.
% However, even if an edge is only seen during testing. Maybe the model has observed the associated nodes extensively during training. TET plots do not allow visualizing the proportion of nodes that the model trains one. 

\newcommand{\Htrain}{\Hcal_{train}}
\newcommand{\Htest}{\Hcal_{test}}

\section{The Impact of Negative Sampling on DLP Evaluation\label{sec:neg_sampling}}
In this section we explore in detail the impact of the negative sampling strategy on the DLP task.

\subsection{Node and Edge-level Categories for Time-based Partitioning\label{sec:taxonomy}}
\newcommand{\powk}[1]{#1^{(k)}}

In Section \ref{background}, we emphasized the importance of comparing scores for positive and associated negative edges when evaluating the predictive abilities of a DLP algorithm. The selection of these negative edges significantly impacts AUC/MRR/AP scores.
To address the challenge of negative edge selection, a naive random approach may uniformly choose negative nodes $\powk{u}$ and $\powk{v}$ from $\Ucal$. However, this often results in overly easy negative edges due to the low likelihood of most node pairs interacting at time $t$\cite{poursafaeiBetterEvaluationDynamic2023}.
A more targeted sampling strategy can avoid this issue of easy negative samples. 
Poursafaei et al. \cite{poursafaeiBetterEvaluationDynamic2023} propose to distinguish edges that appear only in the training set (\textbf{Historical Edges; HE}), only in the test set (\textbf{Inductive Edges; IE}), or both. This approach generates more realistic and challenging negative events.

Building further on this, we note that predicting for a node never seen during training (\textbf{Inductive Node; IN}) is potentially more difficult than predicting for a node that is observed during training (\textbf{Historical Node; HN}). Indeed, for any HN, the model has direct access to relational behavior during training, for instance enabling it to effectively position the node in a latent space. In contrast, for an IN, the predictions will typically be solely based on 
\begin{enumerate*}
  \item connectivity information between $\Hcal([t_{cut},t])$ and 
  \item eventual node features,
\end{enumerate*}  
making it a distinct challenge.

To summarize, a given edge $(u,v)$ sampled at random can be categorized in different ways, as bundled below in Table \ref{tab:ns_taxonomy}.

\begin{table}[!]
	\centering
	\begin{tabular}{|l|p{3.3cm}|p{5.3cm}|}
		\hline
		\textbf{Category Type} & \textbf{Edge category} & \textbf{Description of edge $(u,v)$} \\
		\hline
		\multirow{1}{*}{General} & Never Observed & $(u,v)$ never appears in $\Hcal$\\
		\hline
		\multirow{3}{*}{Edge-Level} & Historical Edges & $(u,v)$ appears in $\Hcal_{train}$ but never in $\Htest$\\
		\cline{2-3}
		& Inductive Edges &  $(u,v)$ appears in $\Hcal_{test}$ but never in $\Htrain$\\
		\cline{2-3}
		& Overlap Edges & $(u,v)$ appears in both $\Hcal_{train}$ and $\Hcal_{test}$\\
		\hline
		\multirow{6}{*}{\multirowsetup{\centering} Node-Level} & Historical-to-Historical & $u$ and $v$ both appear in $\Hcal_{train}$ but never in $\Hcal_{test}$\\
		\cline{2-3}
		& Historical-to-Inductive & $u$ appears in $\Hcal_{train}$ but never in $\Hcal_{test}$, and $v$ appears in $\Hcal_{test}$ but never in $\Hcal_{train}$\\
		\cline{2-3}
		& Inductive-to-Inductive & $u$ and $v$ both appear in $\Hcal_{test}$ but never in $\Hcal_{train}$\\
		\cline{2-3}
		& Overlap-to-Overlap & $u$ and $v$ both appear in $\Hcal_{train}$ and $\Hcal_{test}$\\
		\cline{2-3}
		& Overlap-to-Historical & $u$ appears in $\Hcal_{train}$ and $\Hcal_{test}$, and $v$ appears in $\Hcal_{train}$ but never in $\Hcal_{test}$\\
		\cline{2-3}
		& Overlap-to-Inductive & $u$ appears in $\Hcal_{train}$ and $\Hcal_{test}$, and $v$ appears in $\Hcal_{test}$ but never in $\Hcal_{train}$\\
		\hline
	\end{tabular}
	\caption{A taxonomy of the possible negative edges $(u,v)$.
		Comparing the score of a positive event $s(u,v,t|\Hcalt)$ against the score $s(u',v',t|\Hcalt)$ of an edge belonging to either of these categories conveys different cues about the skills of the DLP algorithm.\label{tab:ns_taxonomy}}
\end{table}

\begin{table}[t]
  \centering
\begin{tabular}{llrrr}
    \toprule
     &  & TGN & EdgeBank & Preferential  \\
    Dataset & Negative Sampler &  &  &Attachment  \\
    \midrule
    \multirow[c]{5}{*}{uci} & Destination & \bfseries 0.882 & 0.547 & 0.503 \\
     & Historical Edge & \bfseries 0.800 & 0.054 & 0.161 \\
     & Inductive Edge & \bfseries 0.626 & 0.554 & 0.544 \\
     & Historical Destination & \bfseries 0.883 & 0.545 & 0.437 \\
     & Inductive Destination & \bfseries 0.875 & 0.554 & 0.720 \\
    \cline{1-2}
    \multirow[c]{5}{*}{wikipedia} & Destination & \bfseries 0.972 & 0.803 & 0.833 \\
     & Historical Edge & \bfseries 0.790 & 0.303 & 0.477 \\
     & Inductive Edge & 0.663 & \bfseries 0.803 & 0.693 \\
     & Historical Destination & \bfseries 0.894 & 0.799 & 0.440 \\
     & Inductive Destination & \bfseries 0.908 & 0.803 & 0.879 \\
    \cline{1-2}
    \multirow[c]{5}{*}{enron} & Destination & \bfseries 0.812 & 0.707 & 0.527 \\
     & Historical Edge & \bfseries 0.825 & 0.258 & 0.338 \\
     & Inductive Edge & 0.671 & \bfseries 0.758 & 0.688 \\
     & Historical Destination & \bfseries 0.813 & 0.706 & 0.502 \\
     & Inductive Destination & 0.826 & 0.758 & \bfseries 0.951 \\
    % \cline{1-2}
    \bottomrule
\end{tabular}
      
  \caption{
    AUC Scores obtained by the 3 methods on 3 different datasets, using 5 different negative sampling strategies.
    % If we treat DLP as a binary classification problem and evaluate it solely using the AUC, we observe that the performance of TGN is highly dependent on the negative sampling strategy.
    In this example we see that while the method performs well in general compared to heuristic baselines, it can get outperformed by a simple heuristic when evaluated on a specific type of negative samples.
    }
    \label{tab:auc_results}
\end{table}

\subsection{Impact of Negative Sampling: An Empirical Evaluation\label{sec:ns_evaluation}}
In order to illustrate concretely the impact of selecting different types of negative edges on the prediction performance, we conduct a follow-up numerical experiment using TGN, a state-of-the-art Dynamic Link Prediction method.
We consider the TGN-attn method, an instance of the Temporal Graph Network \cite{Rossi_Chamberlain_Frasca_Eynard_Monti_Bronstein_2020} that combines graph attention and gated recurrent units into a single temporal graph embedding method. We consider 3 datasets extracted from the recent benchmark \cite{poursafaeiBetterEvaluationDynamic2023}, and compare TGN-Attn with the EdgeBank and PA baselines, against 5 types of negative samples.

The resulting comparison is shown in Table \ref{tab:auc_results}. 
As expected, in certain instances, negative samples are easily identifiable since even basic heuristics yield very high scores. For example, this is the case with PA and Inductive Destination negative samples. In this scenario, these negative samples consistently receive a score of 0 because, by definition, the destination node linked to the negative event does not appear during the testing phase. As a result, the score for positive events is always equal to or greater than that of negative events.
On the other hand, TGN attempts to infer information about inductive destination nodes by employing a combination of message passing and temporal attention. While this Node Inductivity feature is beneficial, in the Enron Dataset, it may lead to an overestimation of event scores involving these negative destination nodes.

\section{Localizing DLP Errors using Longitudinal Plots\label{sec:long_plots}}
As we have seen, the performance of DLP algorithms can vary drastically depending on the type of negative samples considered. Moreover, the prevalence of each type of negative samples varies over time. In this section we propose a way to visualize this temporal variation.
\subsection{Temporal Edge and Node Activity plots\label{sec:visualizations}}

A \textbf{Temporal Node Activity Plot}  for a CTDG $\Hcal
=\{(u_m,v_m,t_m|m=1,\ldots,M)\}$ is 2-dimensional scatter plot representing for each node the timestamps where it is active.
Each node's activity gets represented by a horizontal line with a specific coordinate $y(u)$. The y-coordinates $\{y(u|u\in\Ucal)\}$ are determined by ranking the nodes based on their first arrival time, breaking ties with last arrival time. 
For each event $(u,v,t)$, points are plotted at coordinates $(t,y(u))$ and $(t, y(u))$, representing the temporal activity of each node as a \emph{horizontal sequence}, or \emph{row} of points.

Similarly, for each edge $u,v$ in the network, an \emph{arrival rank} can be derived by sorting the edges $u,v$ in terms of their first interaction times, and breaking the ties using their last interaction time. A \textbf{Temporal Edge Activity Plot} can then be obtained by plotting each event $(u,v,t)$ associated with the edge $u,v$ at the coordinate $(t, y(u,v))$. The activity of each unique edge $(u,v)$ can thus be visualized as a horizontal sequence of points at the y-coordinate $y(u,v)$.

\subsection{Competing Events View\label{sec:competing_eval}}
From our previous discussion on the impact of negative sampling on aggregated measures such as the AUC, we understand the need to explore the measure and visualize the performance across different data segments.
More specifically, it seems sensible to try understand first how the positive event ranks against a carefully selected set of negative samples (e.g., Historical Destination Node), depending on which node or edge it involves.

\begin{figure}[!]
  \centering
  \begin{subfigure}{0.9\textwidth}
    \includegraphics[width=\textwidth]{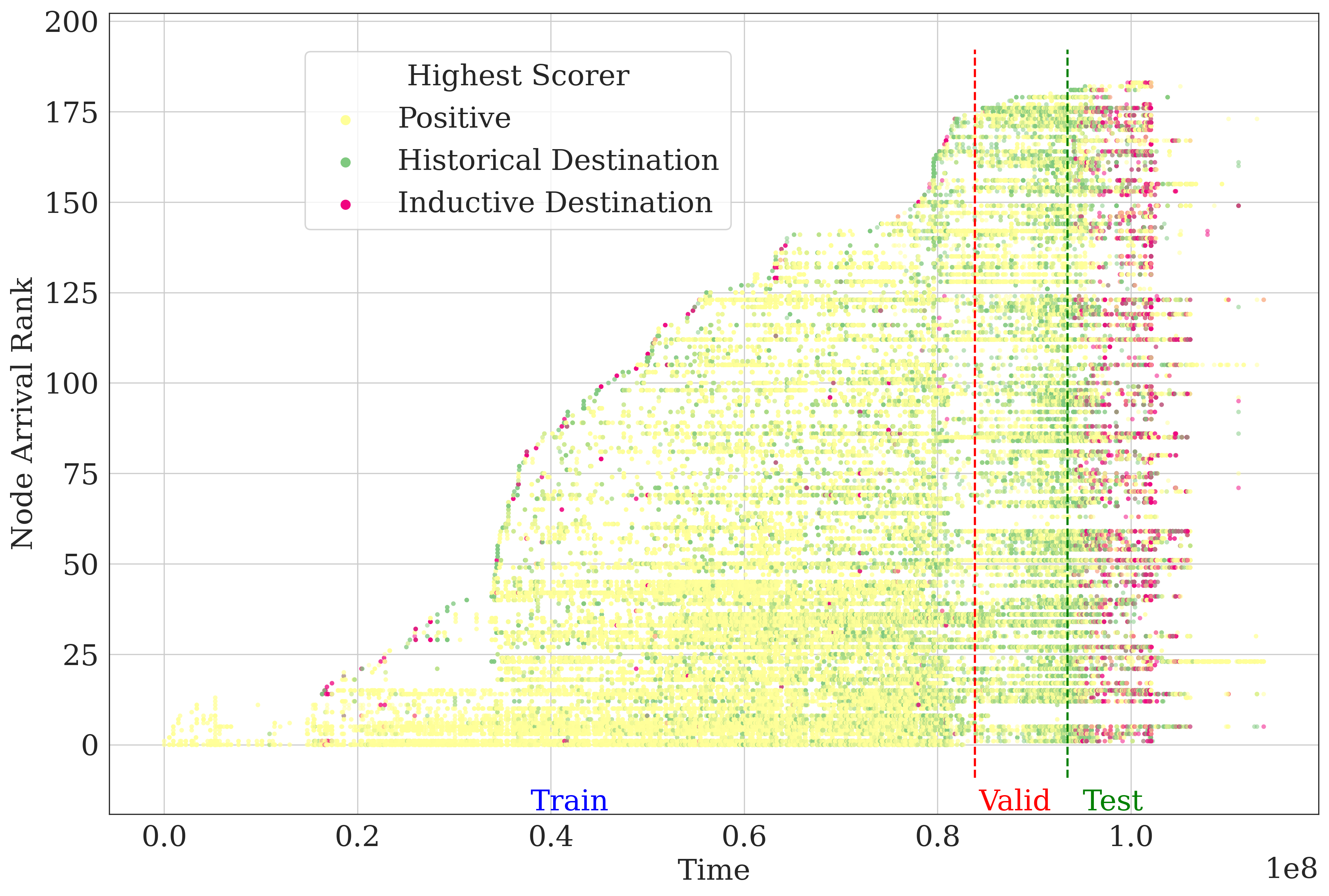}
    \caption{Map of the Predictive error of TGN on a Temporal Node Activity Plot}  
    \label{fig:tnt_ns_eval}
  \end{subfigure}
  \begin{subfigure}{0.9\textwidth}
    \includegraphics[width=\textwidth]{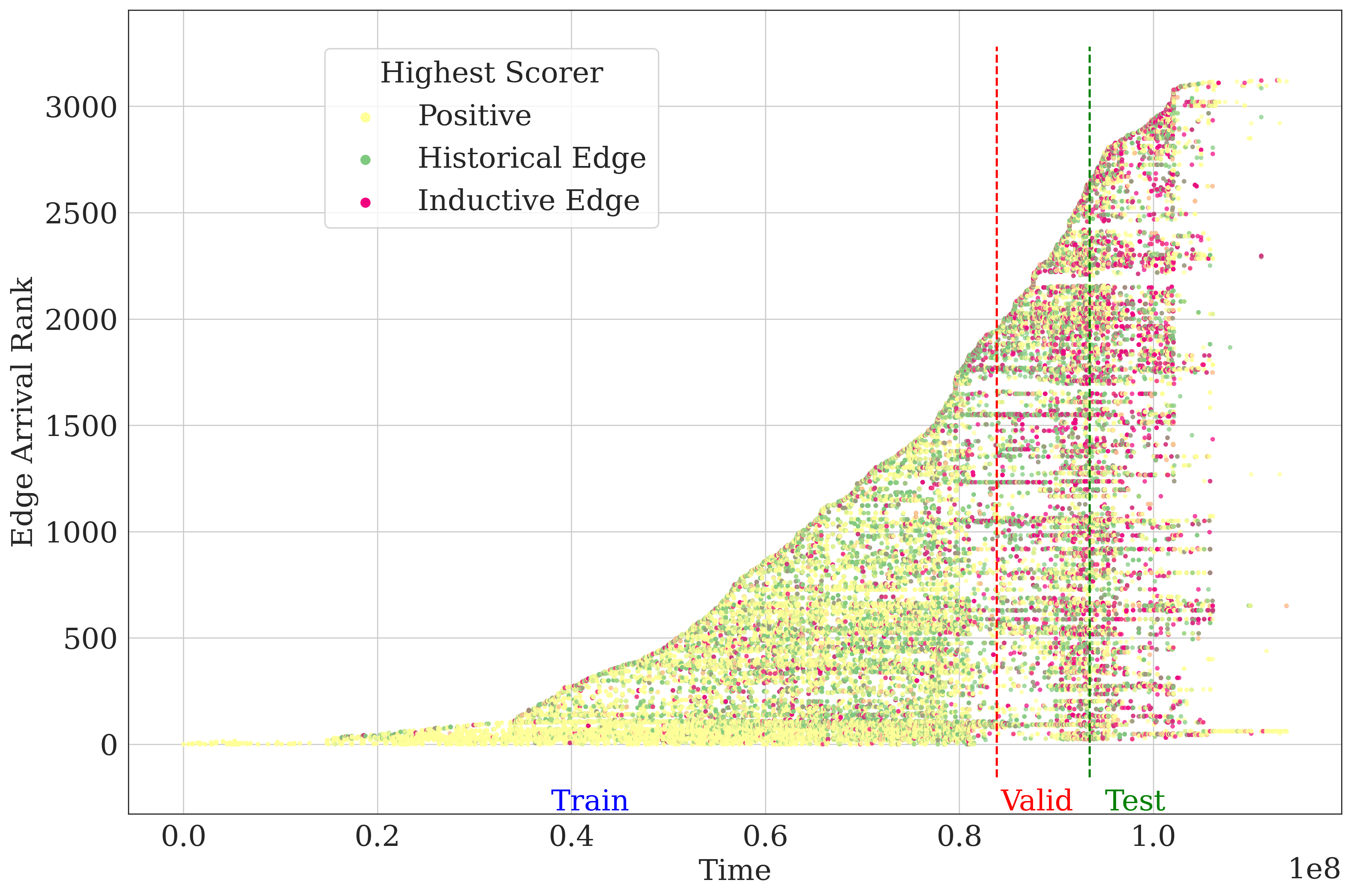}
    \caption{Map of the Predictive error of TGN on a Temporal Edge Activity Plot}  
  \end{subfigure}
  \caption{
    Visualizing the event with the highest score among the set of (Positive, Historical Edge/Destination, and Negative Edge/Destination) over time.
Interestingly, on Figure \ref{fig:tnt_ns_eval}, it can be seen that, during the training period, the model scores the event link with an Inductive Destination node higher than the positive event. This finding supports the notion that the model requires a `warm-up' set of interactions for each node before accurately predicting its interactions.
  }
  \label{fig:ns_eval}
\end{figure}

To further explore this question, we propose to view the task as a competition between various event types, including the positive event. Each event is then labeled and colored based on which candidate attains the best score.
As can be observed in Figure \ref{fig:ns_eval}, when we sample negative events $(u',v',t)$ such that $(u',v')$ appears \emph{exclusively} in the test set, the model readily distinguishes these negative events as negative during the training period, but exhibits more errors during the testing period.

\begin{figure}[!]
  \centering
  \begin{subfigure}{0.9\textwidth}
    \includegraphics[width=\textwidth]{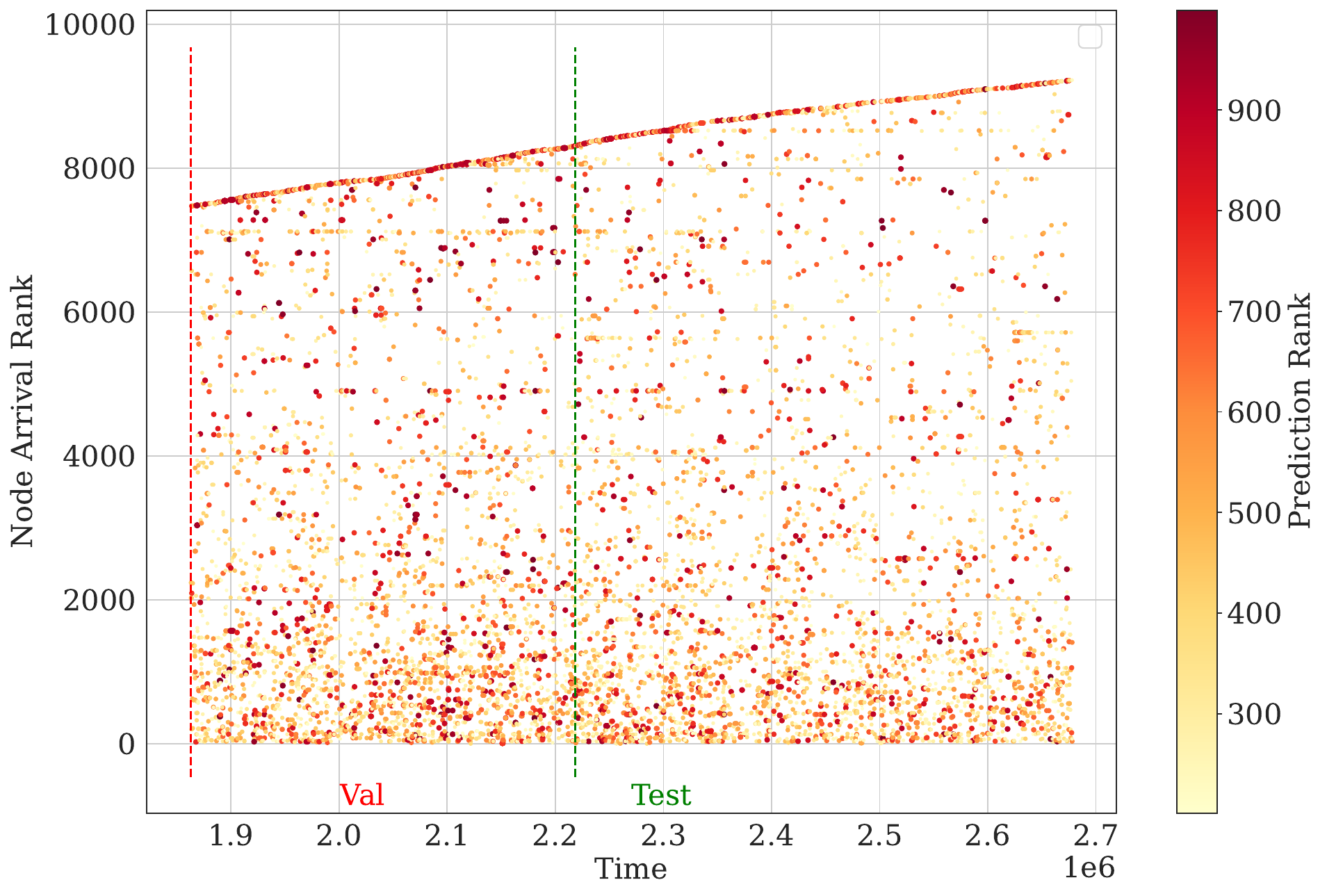}
    \caption{Prediction Rank of events visualized on a Temporal Node Activity plot}
  \end{subfigure}
  \begin{subfigure}{0.9\textwidth}
    \includegraphics[width=1.14\textwidth]{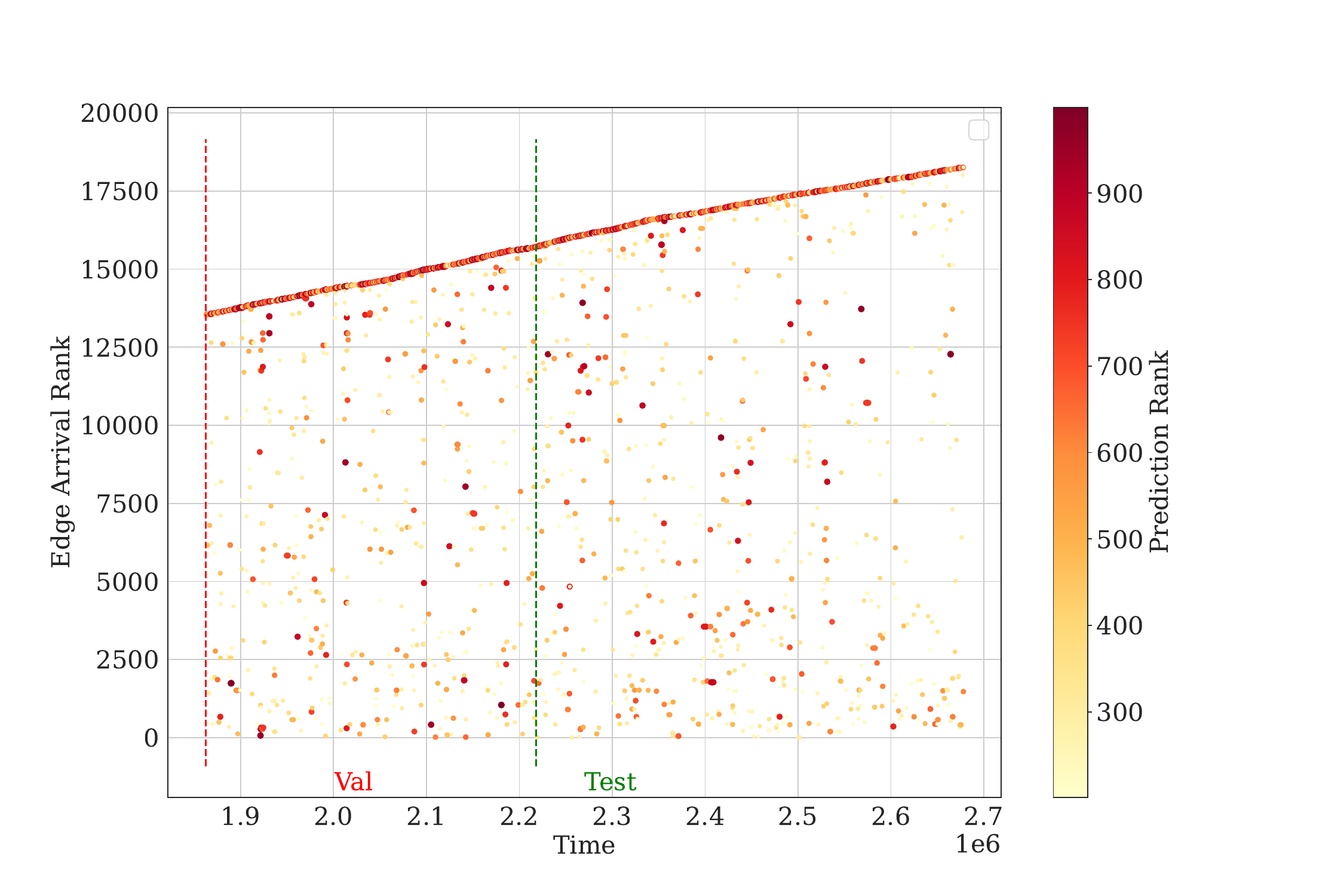}
    \caption{Prediction Rank of events visualized on a Temporal Edge Activity plot}
  \end{subfigure}
  \caption{Longitudinal maps of the prediction rank of each \emph{true} event $(u,v,t)$ among the list of $K$ negative events $(u^{(k)}, v^{(k)}, t)$. 
  Only the events having a prediction rank higher than 200 are shown.}
  \label{fig:tnt_tet_wikipedia}
\end{figure}

\begin{figure}[!]
  \begin{subfigure}{0.46\textwidth}
    \includegraphics[width=\textwidth]{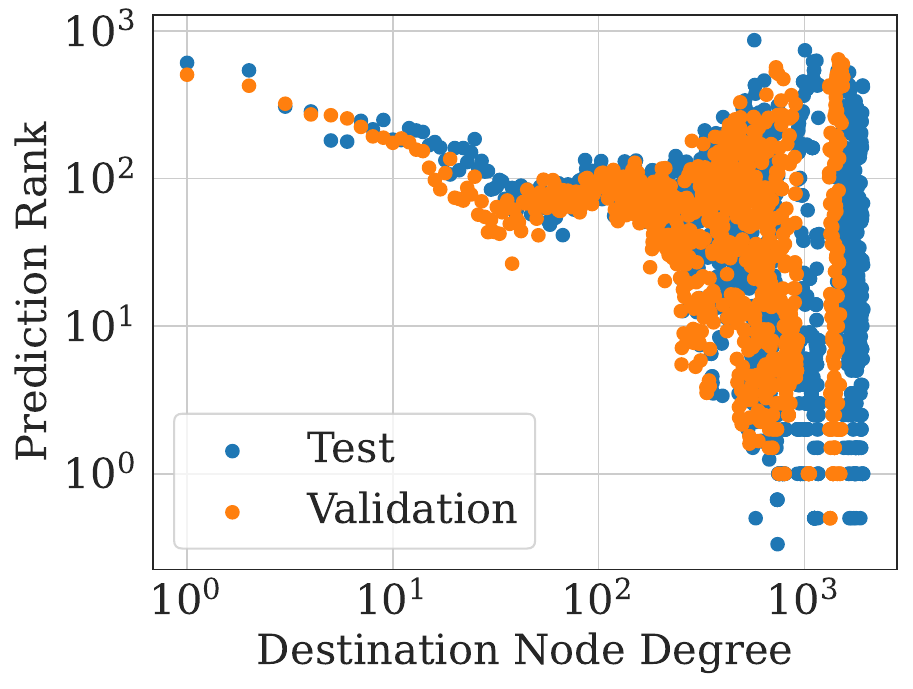}
    \caption{Prediction rank vs Node Degree}
    \label{fig:pred_vs_dst_deg}
  \end{subfigure}
  \begin{subfigure}{0.46\textwidth}
    \includegraphics[width=\textwidth]{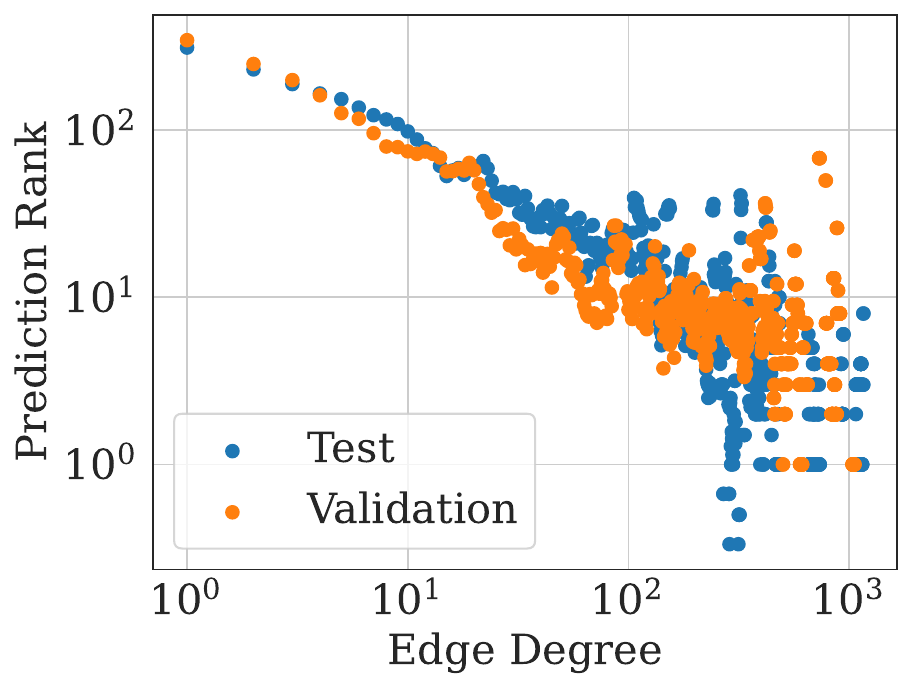}
    \caption{Prediction rank vs. Edge Degree}
    \label{fig:pred_vs_edge_deg}
  \end{subfigure}
  \caption{
     Plot of the prediction rank of the positive event $(u,v,t)$ as a function of (a) the degree of the destination node $v$ or (b) the edge degree $(u,v)$.
  }
  \label{fig:degree}
\end{figure}
\subsection{Visualizing the Prediction Rank of Each Event\label{sec:rank_eval}}

Treating DLP as a binary classification leads to now well-understood flaws, resulting often in unfaithful reporting of model performance, recent efforts have proposed to instead treat the task as a ranking problem.
In this context, for instance, new benchmark datasets such as the ones proposed by Huang et al. \cite{huangTemporalGraphBenchmark2023} for edge-level prediction tasks, include a carefully selected set of $K=1000$ negative events for each positive event.
Mean-Reciprocal Rank is then used as a performance measure.
Following the philosophy of the previous sub-section, we argue that splitting this aggregated measure at the event level and visualize it exactly on Longitudinal plots can lead to various insights.

Thus, we consider the TGBL-Wiki Dataset \cite{huangTemporalGraphBenchmark2023}, splitted using the 70-15-15 proportions, and plot the rank of each event on the TNA and TEA plots on Figure \ref{fig:tnt_tet_wikipedia}. 
On both the node-level and edge-level visualization, we observe that events corresponding to previously unobserved nodes/edges typically have a higher rank than events that are further in the list of events involving a given node/edge.  
We further validate this observation by plotting the prediction rank of the positive event $(u,v,t)$ as a function of the degree of the destination node $v$ (Figure \ref{fig:pred_vs_dst_deg}) or of the edge degree $(u,v)$ (Figure \ref{fig:pred_vs_edge_deg}).
On average, the rank of the positive event is higher for nodes/edges that have been previously unobserved (i.e., possess a low degree at the time of prediction). This observation aligns with the intuition that, for newly encountered entities, the model has limited information about them, making it more prone to misinterpreting a negative event as a positive one.

\section{Related Work}

Link Prediction is a field with a consistent body of research. Seminal work includes \cite{LibenLP2007,lichtenwalterNewPerspectivesMethods2010,luLinkPredictionComplex2011,yangEvaluatingLinkPrediction2015}. However, link prediction in CTDGs come with specific datasets and challenges, that are still underexplored, as the recent TGB challenge \cite{huangTemporalGraphBenchmark2023} indicates.
Methods for predicting links in (discrete or continuous) dynamic graphs have attracted substantial interest in recent years, as surveyed by Kazemi et al.  \cite{kazemiRepresentationLearningDynamic2020a}. 
Competitive methods for DLP include TGN-Attn \cite{Rossi_Chamberlain_Frasca_Eynard_Monti_Bronstein_2020} that we consider in this paper, but also \cite{trivediDYREPLEARNINGREPRESENTATIONS2019,wangInductiveRepresentationLearning2020}.
In this paper,  our aim is not to propose a new method, but rather to introduce a new performance visualization method that can be used to improve existing methods or design new methods.

In terms of the visualization methods considered here, we note that the Temporal Edge Activity plots proposed in Section \ref{sec:long_plots} are closely related to the Temporal Edge Traffic plots from \cite{poursafaeiBetterEvaluationDynamic2023}. However, while in the latter the authors considered exclusively edge-level (`1-st order') visualizations of the network, we complement this by introducing node-level (`0-th order') visualizations.
Node Activity Maps \cite{linharesDyNetVisSystemVisualization2017a} are also notably very similar to Temporal Node Activity plots. However, they rely on a time aggregation of the events, and have not been yet employed to visualize prediction performances of DLP algorithms.

\section{Conclusion}
Recent academic efforts have been dedicated to standardizing Dynamic Link Prediction (DLP) as a machine learning task. The goal is to equip the task with its own evaluation pipelines, baseline methods, and benchmark datasets. However, as a consequence of the high-dimensionality of the data and its non-independent, non-identically distributed nature, deriving a single consistent model validation has encountered numerous challenges. Central to these challenges is negative sampling. In this paper, we assert that the uneven error distribution across nodes/edges and time, might be at the source of these challenges and provide concrete visual cues to support this claim.

Future work in this direction would involve comparing the performance visualizations obtained through other methods. Moreover, plotting the activity map of higher-order entities (e.g., cliques or triangles) within the network could also aid in assessing how performance changes in different parts of the network.

\section*{Acknowledgements}
The research leading to these results has received funding from the European Research Council under the European Union's Seventh Framework Programme (FP7/2007-2013) (ERC Grant Agreement no. 615517), and under the European Union’s Horizon 2020 research and innovation programme (ERC Grant Agreement no. 963924), from the Special Research Fund (BOF) of Ghent University (BOF20/IBF/117), from the Flemish Government under the ``Onderzoeksprogramma Artificiële Intelligentie (AI) Vlaanderen'' programme, and from the FWO (project no. G0F9816N, 3G042220).
% \clearpage

\bibliographystyle{
	% Plot the references as numbers
	abbrvnat
}
\bibliography{References}  %%% Uncomment this line and comment out the ``thebibliography'' section below to use the external .bib file (using bibtex) .





% \end{thebibliography}

\end{document}